# Micromachining & FBG fabrication using point by point technique utilizing femto-second laser

Abu Farzan Mitul, Bohan Zhou, Huiyu Zhao and Ming Han

Electrical & Computer Engineering Department

Michigan State University

Abstract:

Fiber bragg gratings (FBG) has wide variety of applications in sensor and laser devices. In this work, we have fabricated FBG using point by point (PbP) technique utilizing fs laser. FBGs spectral characteristics were explored through the variation of grating length, fiber holder position, laser power and coated layer. Fs FBG is fabricated on small mode field diameter fiber which shows enhancement of transmission depth to 2.5 dB. The center wavelength shift is observed from 1520 nm to 1568 nm with the change in translational stage speed. The fs FBGs can be utilized in high pressure sensing, high temperature sensing and hazardous environment monitoring purposes.

Introduction:

Fiber Bragg gratings (FBGs) have wide variety of applications in optical fiber high temperature sensing, optical telecommunications and an essential tool in fiber lasers to achieve robust, monolithic cavity and narrow linewidth [1]. FBGs are widely accepted sensing device because of it`s low cost, small size, immunity to electromagnetic interference and passive nature. In recent years, researchers have shown FBGs to detect chemical gas, biological, physical parameters e,g, strain, temperature, liquid level, transverse force, breath and heart rates and torque etc [2]. FBGs are usually fabricated through the exposure of photo-sensitive optical fiber to ultra-violet (UV) radiation by interfering two beams. Fabrication of FBGs using femtosecond laser utilizes ultrafast laser with sub-picosecond pulse duration which provides an inherent three dimensional resolution to fabrication and it directly modifies the glass structure inside the fiber core. The quality of femtosecond FBGs is inferior to UV counterparts, but FBGs can be inscribed onto single silica single mode fiber (SMF) without pre-processing such as hydrogenation, reducing time and cost [3]. The fs laser pulses absorption mechanism allows the the fs laser to modify the refractive index in transparent materials which enhances the functional capabilities of FBG inscription. Here, non-linear absorption makes it possible to modify the non-photosensitive material and inscribe FBGs through fiber plastic coating as it is transparent to IR fs radiation. FBG can be inscribed directly by the use of fs laser without utilizing phase mask. There are two main ways of FBG inscription i,e, (i) point-by-point (PbP) and (ii) core scanning techniques. In this paper, we have utilized PbP technique to fabricate FBGs using fs laser. In PbP technique, each pulse writes one grating pitch. Phase shift in the grating can be obtained by adding a delay to the clock source which controls the laser pulse repetition rate [4]. Here, the grating period is defined as the ratio of translation speed of the fiber to laser pulse repetition rate. In this paper, we have investigated the development of grating structure on cover glasses for micromachining purposes and fabricated 2nd order FBG using (Corning SMF-28) single mode telecommunication fiber and small mode field diameter fiber.

Experimental setup and fabrication of FBG:

Figure 1 shows the experimental setup for 2nd order PbP FBG fabrication. The setup comprises of astrella-usp-1k ($\lambda$ = 800 nm, f = 1 KHz, P = 5 W), ABL1000 Air-Bearing, Direct-Drive Linear Stage ( total travel = 100 mm, straightness and flatness = ±0.4 µm), Objective lens (40X, NA= 0.75), DCU223C CCD camera, fiber holder and 3 axis nanomax piezo actuator (X=Y=Z= 4mm). broadband light source (Thorlabs, CLD1015) and the spectrum analyzer (OSA, Yokogawa AQ6370C, Wavelength range: 600 to 1700nm) with a resolution of .020 nm were employed to analyze the transmission spectrum after stretching the fs FBG. The maximum laser pulse energy (~ 5 mJ) was attenuated by rotating a half waveplate which is incorporated with a linear polarizer as shown in figure 1. PbP FBG fabrication is also analyzed using a telescope system as shown in figure 2 which can provides the capability to change the beam size.

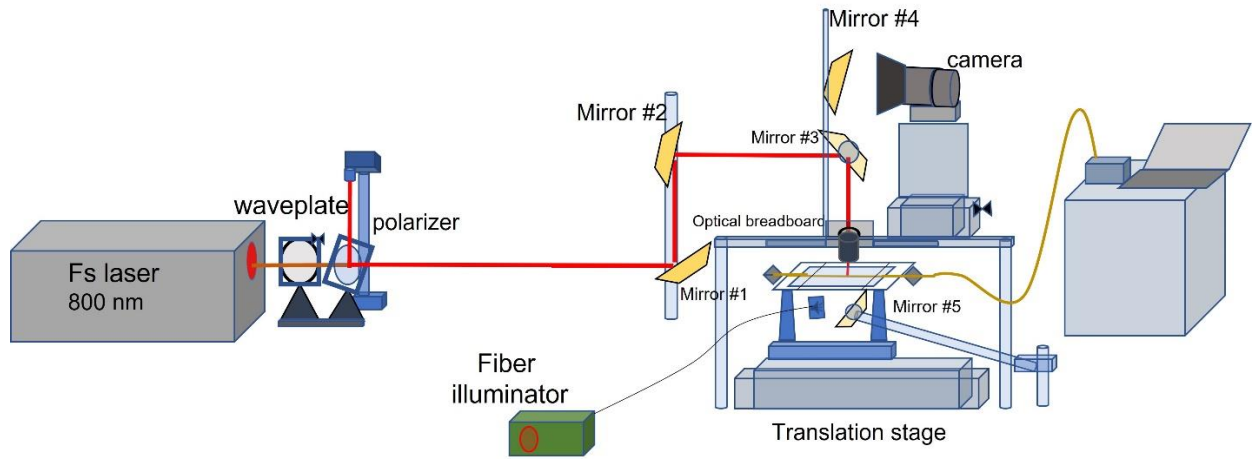

Figure 1: Schematic diagram of experimental setup of PbP fs FBG fabrication

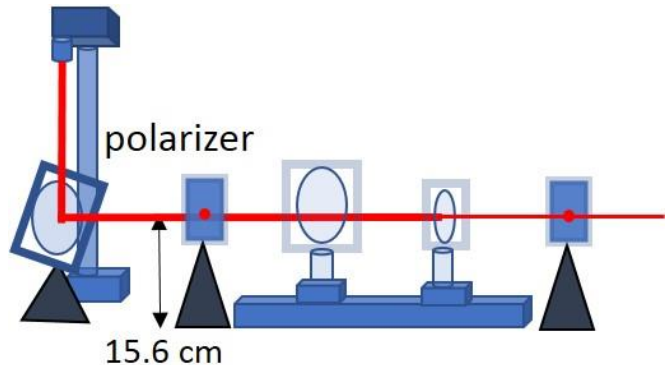

Figure 2: fs beam size controlled using a convex-concave lens combination (telescope system)

One of the advantage of PbP technique in fs FBG fabrication is the use of standard Bragg equation to control/ adjust the Bragg resonance wavelength and reflectivity. Here, the standard Bragg equation is related to the translational stage speed, order of the grating, length and period of the gratings and laser pulse energy [2]. The modified standard Bragg equation is shown below

$$\Lambda = \frac{N \times \lambda_{FBG}}{2n} \quad (1)$$

$\Lambda$ represents the periodicity of the gratings, $\lambda_{FBG}$ is the bragg wavelength, $n$ is the effective refractive index of the waveguide mode, N represents the order of the grating. For first order grating with 1550 nm Bragg wavelength, the periodicity be

$$\Lambda = \frac{\lambda_{FBG}}{2n} = \frac{1550}{2 \times 1.446} = 535.96 \text{ nm (1st order)}$$

Hence, the periodicity for 2nd order is 1071.92 nm. The translational stage speed can be found using equation 2:

$$S_{target} = \Lambda \times f \quad (2)$$

Here, f is the repetition rate of the fs laser and $S_{target}$ is the translational stage speed which relates the periodicity ($\Lambda$), reflection peak wavelength ($\lambda_{FBG}$) and order of gratings (N).

For 2nd order grating, the translational stage speed be ,

$$S_{target} = 2 \times \Lambda \times 1000 = 1071.92 \text{ um/sec} = 1.07 \text{ mm/sec}$$

We tuned the translational stage speed from 1.01 mm/sec to 1.10 mm/sec in order to investigate the wavelength shift, $\lambda_{FBG}$ as shown below.

$$S_{min}(\lambda = 1510\ nm) = 1.044\ mm/sec \text{ and } S_{max}(\lambda = 1590\ nm) = 1.099\ mm/sec$$

Results and discussions:

Fiber core is placed as a sandwiched structure as shown in figure 3. Two dummy fiber is placed on two sides of optical fiber in order to balance the cover glass on top of fiber core. Index matching gel is put on top of fiber core before placing the cover glass. Fiber core has to be placed in a precise manner so that the laser beam strikes the fiber core. The misalignment of fiber core will cause the laser beam to make damages on top of fiber core or, bottom of the fiber core as shown in figure 4. The fiber is placed on a 5D translational stage so that the angular misalignment can also be adjusted.

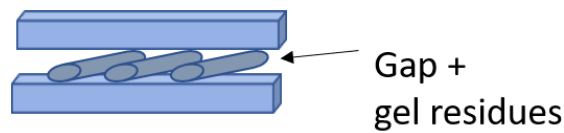

Figure 3: Schematic view of laser beam focusing onto the fiber core

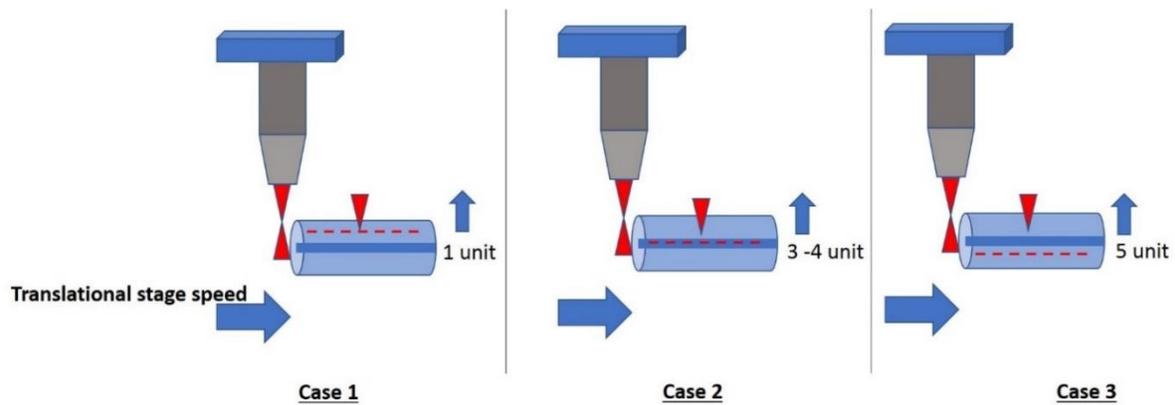

Figure 4: Schematic view of laser beam focusing onto the fiber core

Figure 5(a) shows laser damage points on SM fiber where laser beam was not focused to the fiber core. Figure 5(b) shows laser damage points on fiber core where 5D translational stage was adjusted to position the fiber correctly.

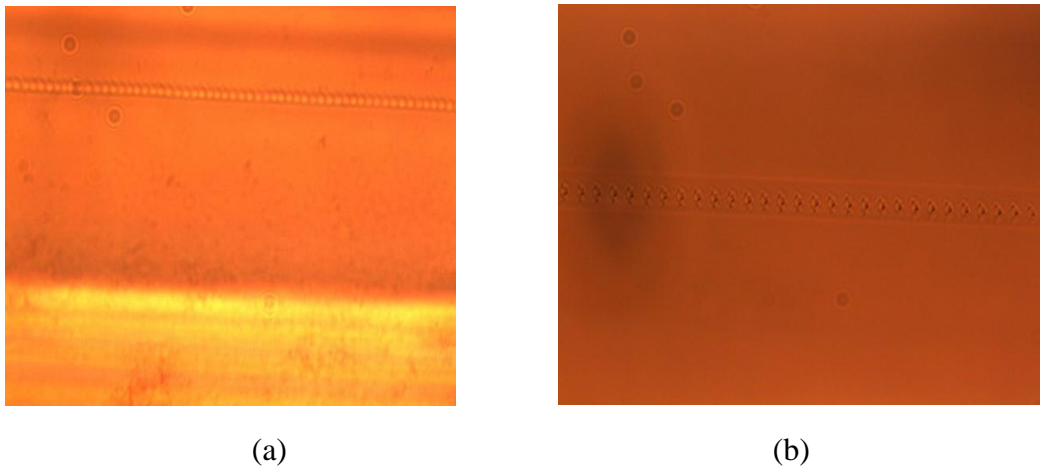

(a)                          (b)

Figure 5: (a) "dot" size damage on SM fiber, (b) damage focused on fiber core

The translational stage is moved at 1.03 mm/sec and laser beam power was fixed to 2 mW. figure 6(a) shows the reflection spectrum of FBG for 6 mm grating length and figure 6(b) shows the dot size structure onto the fiber core. Again, the translational stage is moved at 1.05 mm/sec and figure 7(a) shows the reflection spectrum of FBG for same grating length as previous experiment. Figure 5(b) shows the microscopic image of fs FBG. Due to the change in translational stage speed from 1.03 mm/sec to 1.05 mm/sec, the reflection peak changed from 1537 nm to 1568 nm

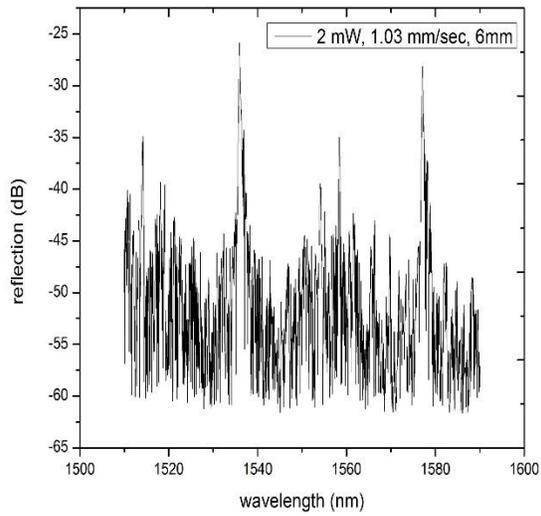 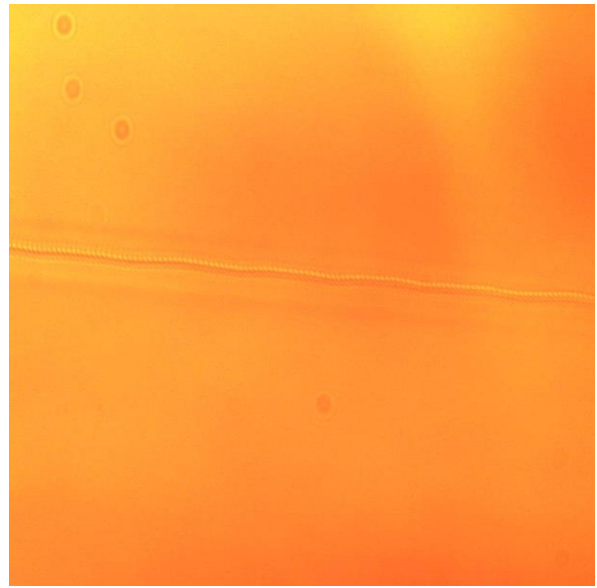

(a) (b)

Figure 6: (a) Microscopic image of fs FBG on fiber core, (b) Reflection spectrum of $2^{nd}$ order FBG translational stage speed of 1.03 mm/sec

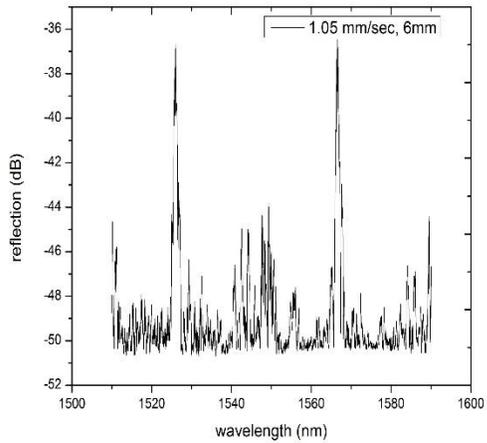 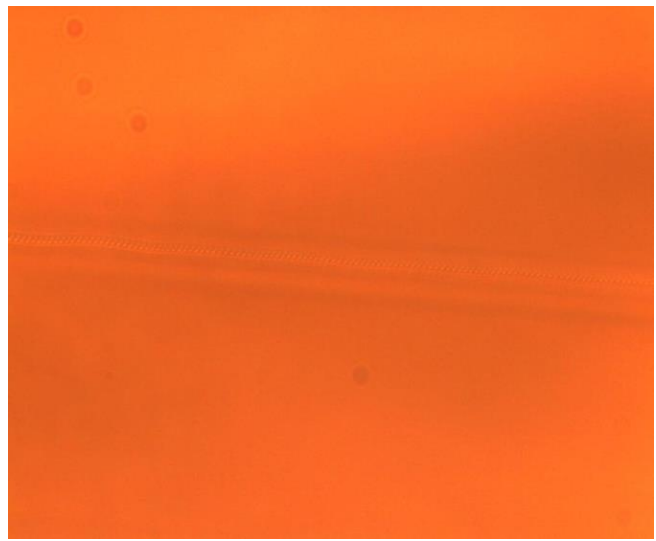

(a) (b)

Figure 7: (a) Microscopic image of fs FBG on fiber core, (b) Reflection spectrum of $2^{nd}$ order FBG with translational stage speed of 1.05 mm/sec

The translational stage is moved to 8 mm where the grating length is maintained of 6mm. Figure 8 shows the schematic view of the setup where laser beam was turned on to create a grating length of 6 mm. The fs FBG is fabricated with coated layer.

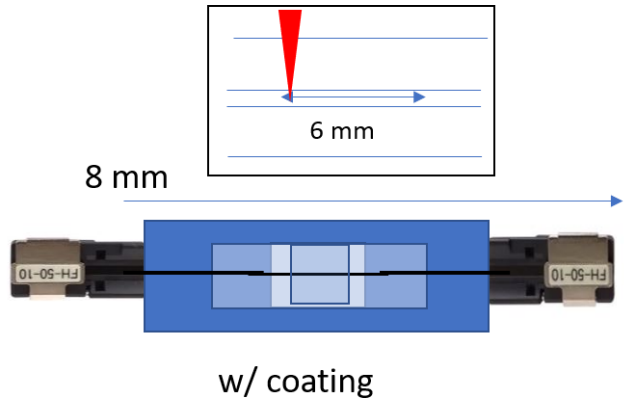

Figure 8: Schematic view of movement of translational stage to 8 mm where the grating length is 6 mm

Figure 9 and 10 show the fs FBG reflection spectrum and microscopic image where the fiber coating was not removed. The translational stage is moved at 1.02 mm/sec and grating length of 8 mm was maintained for figure 9(a) fs FBG. The laser beam power was reduced to 800µW. Figure 10 (a) shows the reflection spectrum of fs FBG where the grating length is 7 mm and translational stage speed is 1.04 mm/sec. Due to the change in translational stage speed from 1.02 mm/sec to 1.04 mm/sec, the reflection spectrum peak changed from 1520 nm to 1551 nm.

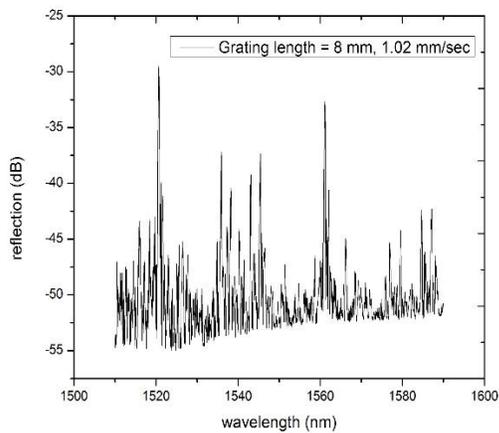
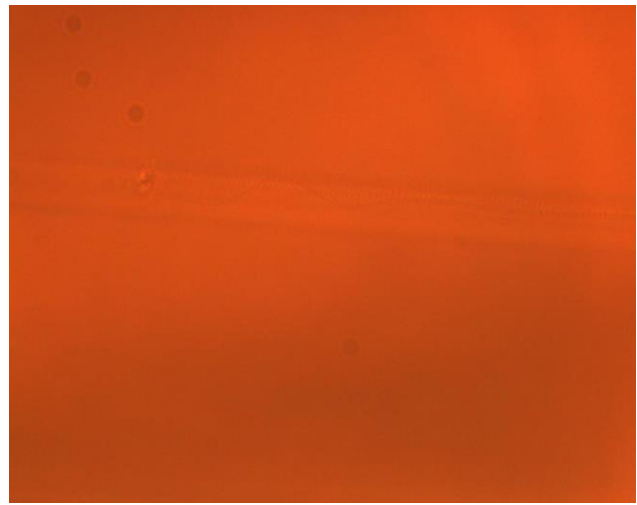

(a)            (b)

Figure 9: (a) Microscopic image of fs FBG on fiber core, (b) Reflection spectrum of 2$^{nd}$ order FBG with translational stage speed of 1.02 mm/sec and grating length 8 mm

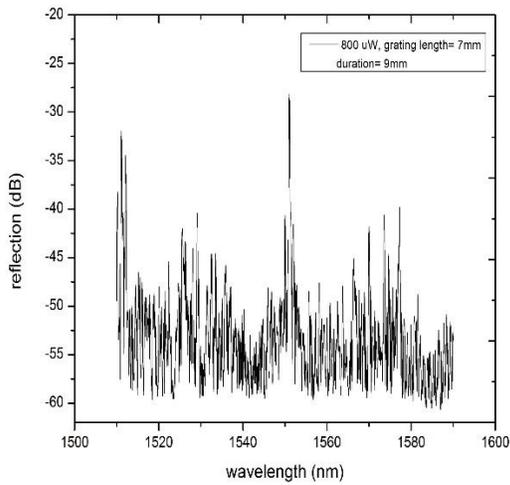 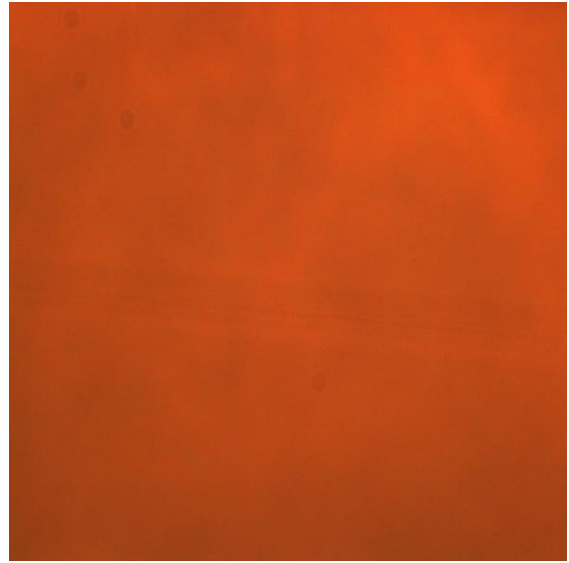

(a)                  (b)

Figure 10: (a) Microscopic image of fs FBG on fiber core, (b) Reflection spectrum of $2^{nd}$ order FBG with translational stage speed of 1.04 mm/sec and grating length 7 mm

Figure 11(a) shows transmission spectrum of $2^{nd}$ order FBG where the translational stage speed is 1.04 mm/sec, grating length of 7 mm and optical power is 800 uW. Two transmission peak was at around 1550 nm and at 1591 nm respectively. The fiber is stretched and the transmission peak shifted towards the higher wavelength as shown in figure 11(b).

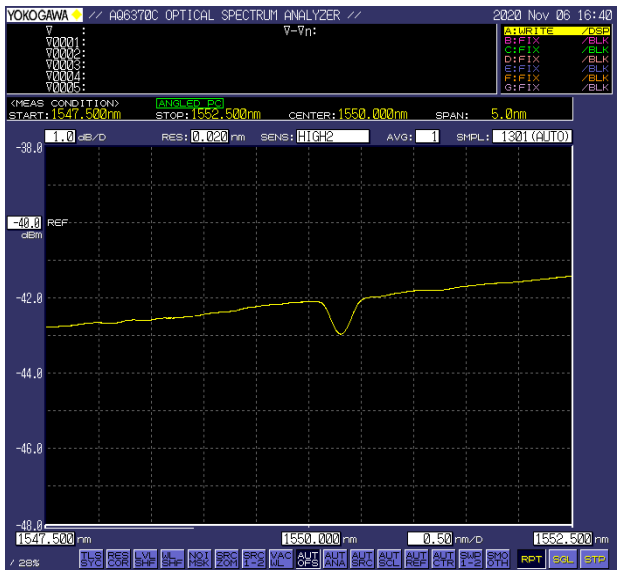 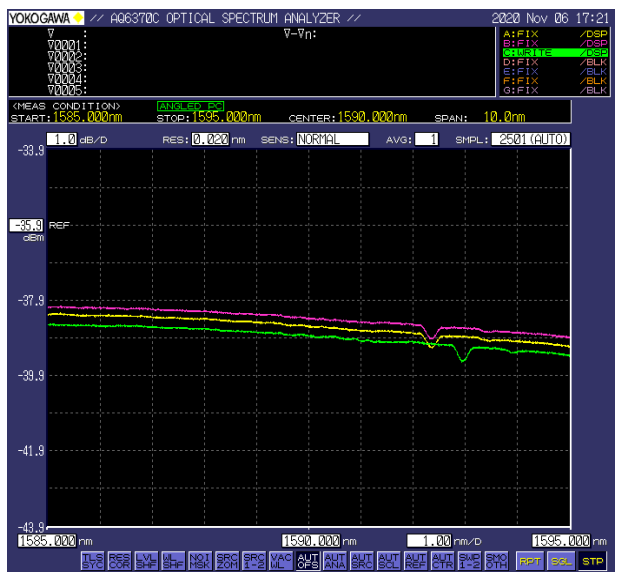

(a)                  (b)

Figure 11: (a) Transmission spectrum of $2^{nd}$ order FBG with translational stage speed of 1.04 mm/sec, grating length 7 mm and optical power of 800 uW, (b) Stretching effect on transmission peak of fabricated fs FBG

Figure 12(a) shows the microscopic image of small mode field diameter fiber where the mode diameter is around 3.3 µm.

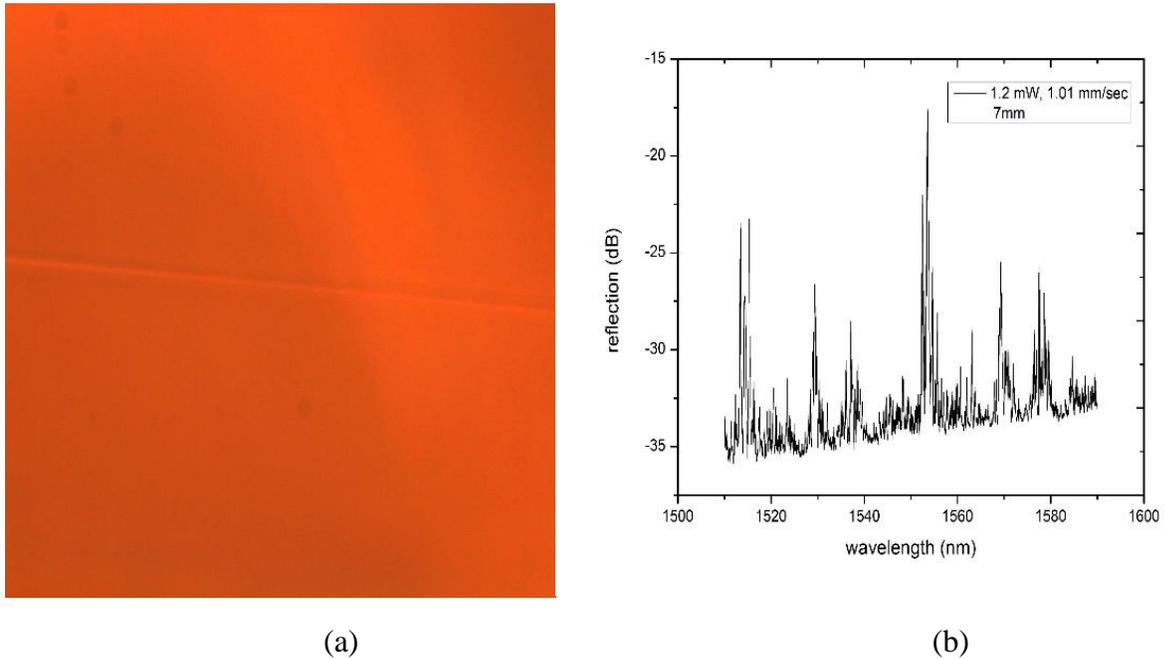

(a)                          (b)

Figure 12: (a) Microscopic image of Small mode field diameter fiber, (b) Reflection spectrum of $2^{nd}$ order FBG with translational stage speed of 1.01 mm/sec and grating length of 7 mm on Small mode field diameter fiber

Figure 12(b) shows the reflection spectrum of fs FBG on small mode field diameter fiber where the translational stage speed is 1.01 mm/sec, grating length of 7 mm and laser beam power was set to 1.2 mW and figure 13 shows the transmission spectrum for the above configuration. In small mode field diameter fiber, the transmission depth increased to 2.5 dB at 1551 nm.

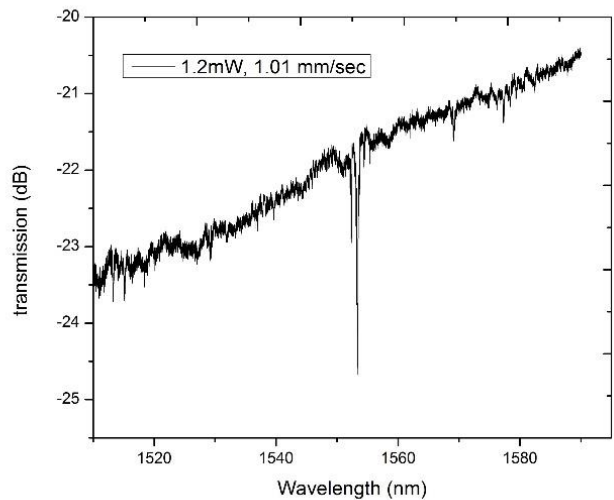

Figure 13: Transmission spectrum of 2nd order FBG with translational stage speed of 1.01 mm/sec and grating length of 7 mm on Small mode field diameter fiber

Conclusion:

2nd order FBG is fabricated using (Corning SMF-28) single mode telecommunication fiber and small mode field diameter fiber. Small mode field diameter fiber (with coating) demonstrated enhancement of transmission depth to 2.5 dB. There was a change in reflection peak wavelength from 1520 nm to 1568 nm as the translational stage moves from 1.02 mm/sec to 1.05 mm/sec. Fabricated fs FBG will be used for high temperature and high stress sensing purposes in near future.